\documentclass[aps,prl,twocolumn,showpacs,byrevtex,superscriptaddress]{revtex4}

 \usepackage{epsfig}
 \usepackage{color}
 \usepackage{amsmath}
 \usepackage{amsthm}
 \usepackage{amsfonts}
 \usepackage{amssymb}
\usepackage{graphicx}
\usepackage{epstopdf}

\def\be{\begin{equation}} \def\eea{\end{eqnarray}}
\def\ee{\end{equation}} \def\bea{\begin{eqnarray}}
\def\ea{\end{array}} \def\ba{\begin{array}}

\newcommand{\bel}[1]{\begin{equation}\label{#1}}

\def\zzz{{\mathchoice {\hbox{$\sf\textstyle Z\kern-0.4em Z$}}
{\hbox{$\sf\scriptstyle Z\kern-0.3em Z$}}
{\hbox{$\sf\scriptscriptstyle Z\kern-0.2em Z$}} {\hbox{$\sf\textstyle
Z\kern-0.4em Z$}}}}

\begin{document}

\title{Observation of Replica Symmetry Breaking in the 1D Anderson Localization Regime in an Erbium-Doped Random Fiber Laser}

\begin{abstract}
The analogue of the paramagnetic to spin-glass phase transition in disordered magnetic systems, leading to the phenomenon of replica symmetry breaking, has been recently demonstrated in a two-dimensional random laser consisting of an organic-based amorphous solid-state thin film. We report here the first demonstration of replica symmetry breaking in a one-dimensional photonic system consisting of an erbium-doped random fiber laser operating in the continuous-wave regime based on a unique random fiber grating system, which plays the role of the random scatterers and operates in the Anderson localization regime. The clear transition from a photonic paramagnetic to a photonic spin glass phase, characterized by the probability distribution function of the Parisi overlap, was verified and characterized. In this unique system, the radiation field interacts only with the gain medium, and the fiber grating, which provides the disordered feedback mechanism, does not interfere with the pump.
\end{abstract}

\author{Anderson~S.~L. Gomes}
\email{anderson@df.ufpe.br}
\affiliation{Departamento de F\'{\i}sica,
Universidade Federal de Pernambuco, 50670-901, Recife-PE, Brazil}

\author{Bismarck C. Lima}
\affiliation{Departamento de F\'{\i}sica,
Universidade Federal de Pernambuco, 50670-901, Recife-PE, Brazil}

\author{Pablo I.~R. Pincheira}
%\email{priquelmepincheira@gmail.com}
\affiliation{Departamento de F\'{\i}sica,
Universidade Federal de Pernambuco, 50670-901, Recife-PE, Brazil}

\author{Andr\'e L. Moura}
%\email{andre.moura@fis.ufal.br}
\affiliation{Departamento de F\'{\i}sica,
Universidade Federal de Pernambuco, 50670-901, Recife-PE, Brazil}
\affiliation{Grupo de F\'{\i}sica da Mat\'eria Condensada, N\'ucleo de Ci\^encias Exatas - NCEx, Campus Arapiraca, Universidade Federal de Alagoas, 57309-005, Arapiraca-AL, Brazil}

\author{Mathieu Gagn\'e}
%\email{priquelmepincheira@gmail.com}
\affiliation{Fabulas Laboratory, Department of Engineering Physics, Department of Electrical Engineering, Polytechnique Montreal, Montreal, H3C 3A7, Canada}

\author{Raman Kashyap}
%\email{priquelmepincheira@gmail.com}
\affiliation{Fabulas Laboratory, Department of Engineering Physics, Department of Electrical Engineering, Polytechnique Montreal, Montreal, H3C 3A7, Canada}

\author{Ernesto~P. Raposo}
%\email{ernesto@df.ufpe.br}
\affiliation{Laborat\'orio de F\'{\i}sica
Te\'orica e Computacional, Departamento de F\'{\i}sica,
Universidade Federal de Pernambuco, 50670-901, Recife-PE, Brazil}

\author{Cid B. de Ara\'ujo}
%\email{cid@df.ufpe.br}
\affiliation{Departamento de F\'{\i}sica,
Universidade Federal de Pernambuco, 50670-901, Recife-PE, Brazil}

\date{\today }


\maketitle

A Random Fiber Laser (RFL), which is the one-dimensional (1D) fiber waveguide version of random lasers (RLs) in bulk materials, was pioneered by de Matos {\it et al.} in 2007~\cite{1}, using a colloidal Rhodamine 6G+TiO$_2$ in the core of a photonic crystal fiber. It was followed by the demonstration of a RFL based on randomly spaced fiber Bragg gratings (FBG) inscribed in an erbium-doped fiber in 2009~\cite{2,3}. Soon after, Turitsyn and co-workers demonstrated a RFL in conventional optical fibers via Rayleigh scattering due to the refractive index fluctuations, amplified through the Raman effect~\cite{4}. While the RFL reported in Ref.~\cite{1} operated in the so-called incoherent feedback regime, Hu {\it et al.}~\cite{5} later demonstrated the operation of a RFL similar to \cite{1}, but with the required properties to operate in the coherent feedback regime. Several other reports and applications of RFL in different types of fibers followed afterwards, as reviewed in Ref.~\cite{6}.

Other 1D or quasi-1D RLs have been reported in the literature, both from the theoretical and experimental point of view~\cite{7,8,9,10,11,12,13,14}. All of these reports describe the lasing mechanism or laser related properties of 1D RLs, with particular emphasis on the threshold behavior.

RLs in general differ from conventional lasers as the feedback is provided by scattering rather than by a set of static mirrors. Such scatters can be passive, and therefore embedded in a gain medium, or active, playing a double role of scattering and gain medium. Examples of the former are dye+TiO$_2$ colloids, which were employed in the first unambiguous experimental demonstration of random lasers~\cite{15}, whereas examples of the latter are ZnO powders~\cite{16} or rare earth doped powders, such as Nd$^{3+}$ in different hosts~\cite{17,18}. Although operating mirrorless, RLs present cavity modes~\cite{19,20}. The optical feedback provided by the scatterers has been classified as incoherent or coherent feedback, although there is no universal agreement. Keeping this notation, when the incoherent feedback dominates, the spectrum above the threshold is smooth, whereas in the case of coherent feedback, it presents several narrow spikes on the top of the smooth background due to amplified spontaneous emission. These spikes are directly related to the modes of the random laser, and were first identified in the work of Ref.~\cite{16}. Further work on random lasers is reviewed in Ref.~\cite{21}, including applications and highlights of interdisciplinary results. Among these, we point out the very recent experimental demonstration of replica symmetry breaking (RSB) in RLs~\cite{22}, a concept inherent to the theory of spin glasses and complex systems~\cite{23} and theoretically proposed to be observed in photonic systems, particularly random lasers, in 2006~\cite{24,25}. In short, the RSB approach describes how identical systems under identical initial conditions can reach different states. By investigating the distribution of correlations between RL intensity fluctuations from pulse-to-pulse, the authors of Refs.~\cite{24,25} defined an analogue to the Parisi overlap order parameter ($q_{\mbox{\scriptsize max}}$) and found evidence of a transition from the photonic paramagnetic to a glassy phase of light -- a photonic spin-glass phase. The experimental demonstration of Ref.~\cite{22} employed a RL with functionalized thiophene-based oligomer commonly named (T$_5$OC$_x$) in amorphous solid state in planar geometry, pumped by a frequency doubled (532 nm) pulsed (10 Hz, 6 ns) Nd:YAG laser. This 2D RL emitted radiation with the several spikes depicting the modes riding on a broad pedestal around 610 nm, when high resolution spectral measurements were employed. When a lower spectral resolution was employed, a somewhat smooth spectra was measured. Using the replica theory, they analyzed the shot-to-shot intensity fluctuation to obtain the RSB signature and clearly demonstrated the photonic paramagnetic to photonic spin-glass phase. The theoretical and related experimental work has been reviewed in detail in Refs.~\cite{26,27,28}.

In the present work, we describe the first experimental results of RSB in a 1D system, operating in the cw regime, employing an erbium-doped RFL (Er-RFL) in which the scatterers are fiber Bragg gratings (FBG) randomly spaced and prepared in a unique way such as to provide a suitable density of scatterers that makes the system operate in the coherent feedback regime, where two modes are clearly observed.

The Er-RFL used in this work was the same device employed in the report of Ref.~\cite{2}, where the fabrication and fiber Bragg grating inscription details can be found. For the understanding of the present work, it suffices to say that a polarization maintaining Er doped fiber was used where a continuous grating with randomly distributed phase errors was written, instead of a random array of gratings, as in \cite{3}. This allowed the number of scatterers and the randomness to be significantly raised. An optimized length of 20 cm was used in this work, and the pump source was a cw laser diode operating at 976 nm. The threshold and laser linewidth of $\approx 3$ mW and 0.1 nm (limited by our system's spectral resolution), respectively, were obtained, as compared to $\approx 3$ mW and 0.5 pm for the same parameters demonstrated in \cite{2}. For 100 mW pump power, a 3 mW output power was measured~\cite{2}. For the results reported here, the maximum pump power employed was 80 mW.  As pointed out in Ref.~\cite{2}, the Er-RFL operates in the Anderson localization regime, of course in 1D, meaning that the transmission follows an exponential decay as in the localization theory, and studied in detail by Shapira and Fischer for a similar random-grating array system in a single mode fiber~\cite{29}. Therefore, our Er-RFL operates in the resonant feedback or coherent regime~\cite{1,5,16,21}.

For the statistical measurements
%used to calculate the probability distribution functions (PDF),
a sequence of 1000 spectra was collected for each pump power. To obtain the set of 1000 spectra, the output of the Er-RFL was directed onto an Acton 300i spectrometer with a liquid-N$_2$ infrared CCD camera and 0.1 nm resolution at 1530 nm.

%%FIG 1

\begin{figure}[center]
\centerline{\epsfig{width=7.45 cm,clip,figure=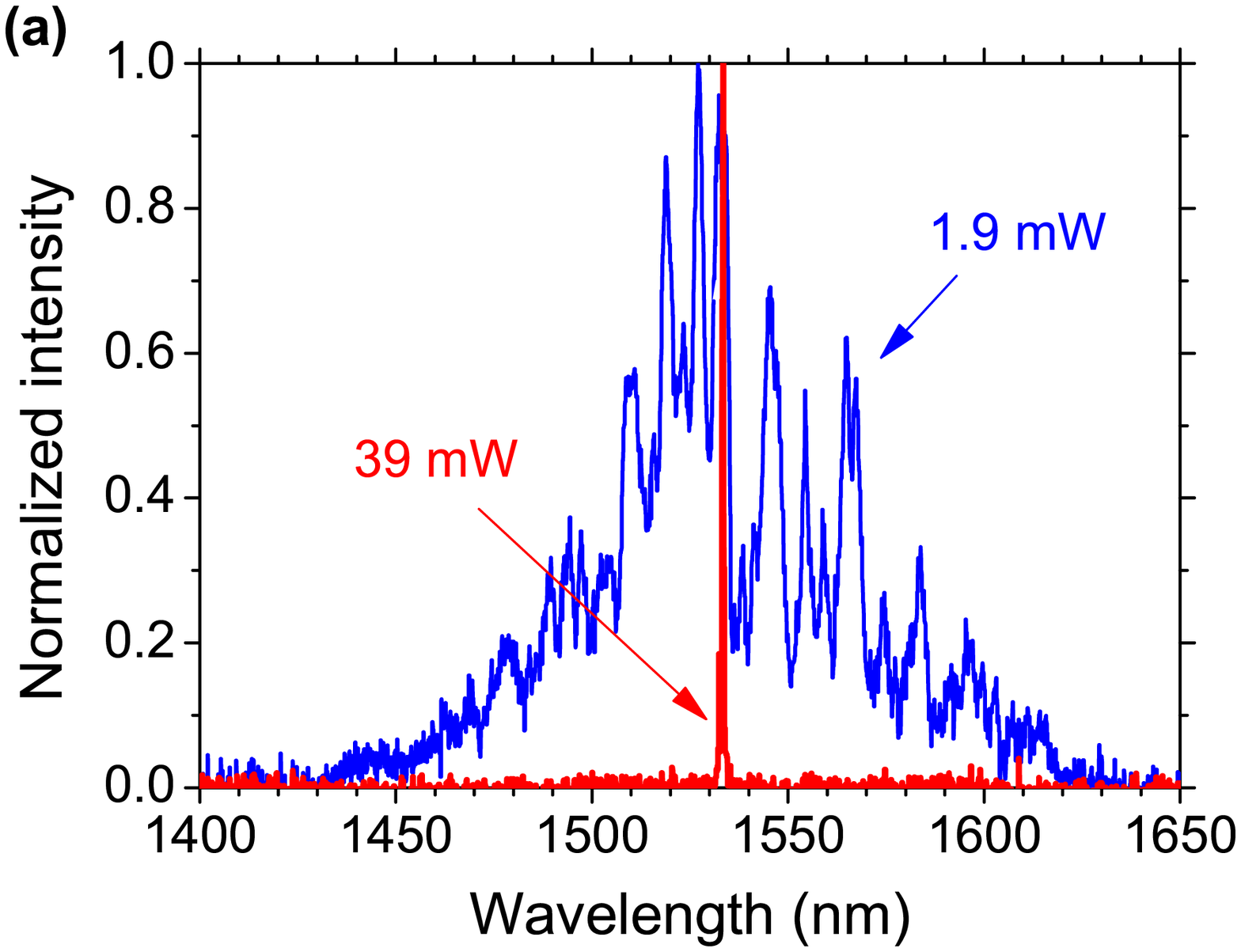}}
%%%\centerline{\epsfig{width=7 cm,clip,figure=Fig.1a_pb_v3.eps}}
%%\centerline{\epsfig{width=7 cm,clip,figure=FigS3.eps}}
%%\centerline{\epsfig{width=10 cm,clip,figure=Fig.1av2.eps}}
%\centerline{\epsfig{width=10 cm,clip,figure=Fig.1b.eps}}
%\centerline{\epsfig{width=11 cm,clip,figure=Fig.1bv3.eps}}
%\hspace{0.46 cm}
\centerline{\epsfig{width=7.45 cm,clip,figure=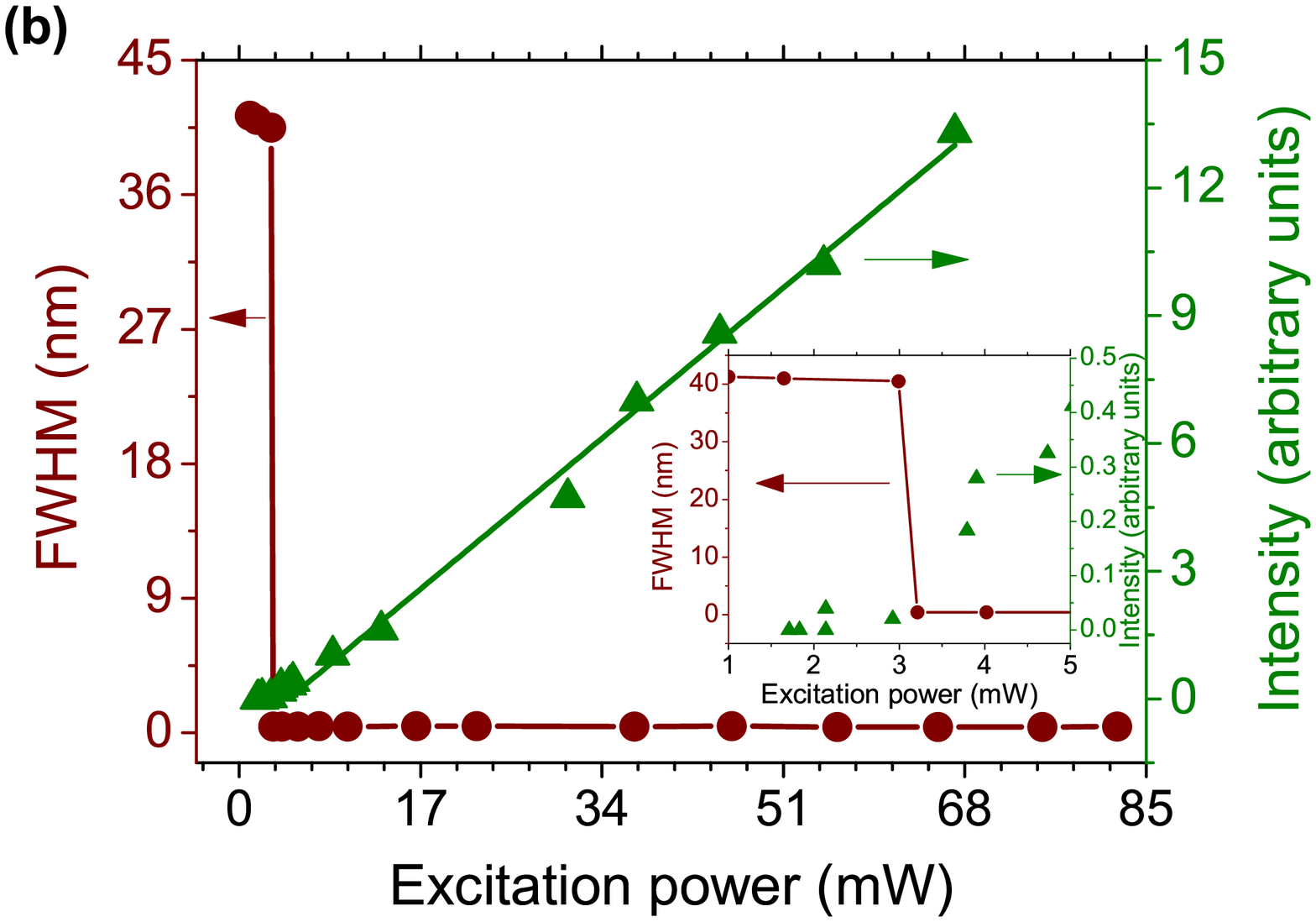}}
%%%\epsfig{width=7.45 cm,clip,figure=Fig.1b_pb_v2.eps}
%%\epsfig{width=10.65 cm,clip,figure=Fig.1bv3.eps}
%\vspace{-0.5cm}
\caption{%{\bf Intensity spectra and the RL threshold.}
(color online).
(a) Output spectrum of Er-RFL at pump powers below (blue curve) and above (red curve) threshold.
(b) FWHM and emitted Er-RFL intensity as a function of pump power.
The inset shows a closer look around the threshold value. The line is a guide to the eyes.}
\label{fig1}
\end{figure}

Figure 1 shows, for the sake of completeness, the Er-RFL characterization. Figure~1(a) shows the spectrum below and above threshold for different pump powers while Fig.~1(b) shows the linewidth reduction and emitted intensity as a function of pump power. The inset gives a closer view around the threshold pump power, where it can be seen that both the emitted intensity and subtle linewidth narrowing occurs at the same pump power around 3mW.

The characterization of the RSB phase transition from the photonic paramagnetic to the spin-glass RL behavior can be quantified by an overlap parameter $q_{\gamma\beta}$ analogue to the Parisi parameter in spin-glass theory~\cite{23}. Two-point correlations can be calculated either among mode amplitudes $a_j$~\cite{25,26,27} or intensities $I_j \propto |a_j|^2$~\cite{22,28}, though the latter are more accessible experimentally. In particular, by measuring fluctuations in the spectral intensity averaged over $N_s$ system replicas, the overlap parameter reads~\cite{22,28}:
\begin{equation}
q_{\gamma \beta} = \frac{ \sum_{k} \Delta_\gamma (k) \Delta_\beta (k)}{ \sqrt{\sum_{k} \Delta^2_\gamma (k)} \sqrt{\sum_{k} \Delta^2_\beta (k)} },
\label{ab}
\end{equation}
where $\gamma$ and $\beta = 1,2,...,N_s$, with $N_s=1000$ in this work, denote the replica labels, the average intensity at the wavelength indexed
by $k$ reads $\bar{I}(k) = \sum_{\gamma = 1}^{N_s} I_\gamma (k)/N_s$,
and the intensity fluctuation is given by $\Delta_\gamma (k) = I_\gamma (k) - \bar{I}(k)$.
In the present context, with a cw laser as the pump source, each emission spectrum collected within a window time of 100 ms is considered a replica, i.e. a copy of the RL system under fairly identical experimental conditions. In order to confirm that the cw measurements were appropriate, we repeated the experiment and, instead of keeping the laser ON all the time, we employed a chopper before the Er-RFL such as to turn the pump beam ON and OFF at 200 Hz. The results were readily reproduced, assuring that the statistical behavior was maintained.
The probability density function (PDF) $P(q)$ describes the distribution of replica overlaps $q=q_{\gamma\beta}$, signaling a photonic uncorrelated paramagnetic or a RSB spin-glass phase if it peaks at $q=0$ (no RSB) or at values $|q| \not = 0$ (RSB), respectively. In this context, the break of replica symmetry means that the most probable value of the intensity correlations between the laser spectra
is no longer null, as in the symmetric case, but instead presents a non-zero value.

\begin{figure}[center]
\centerline{\epsfig{width= 5.1 cm,clip,figure=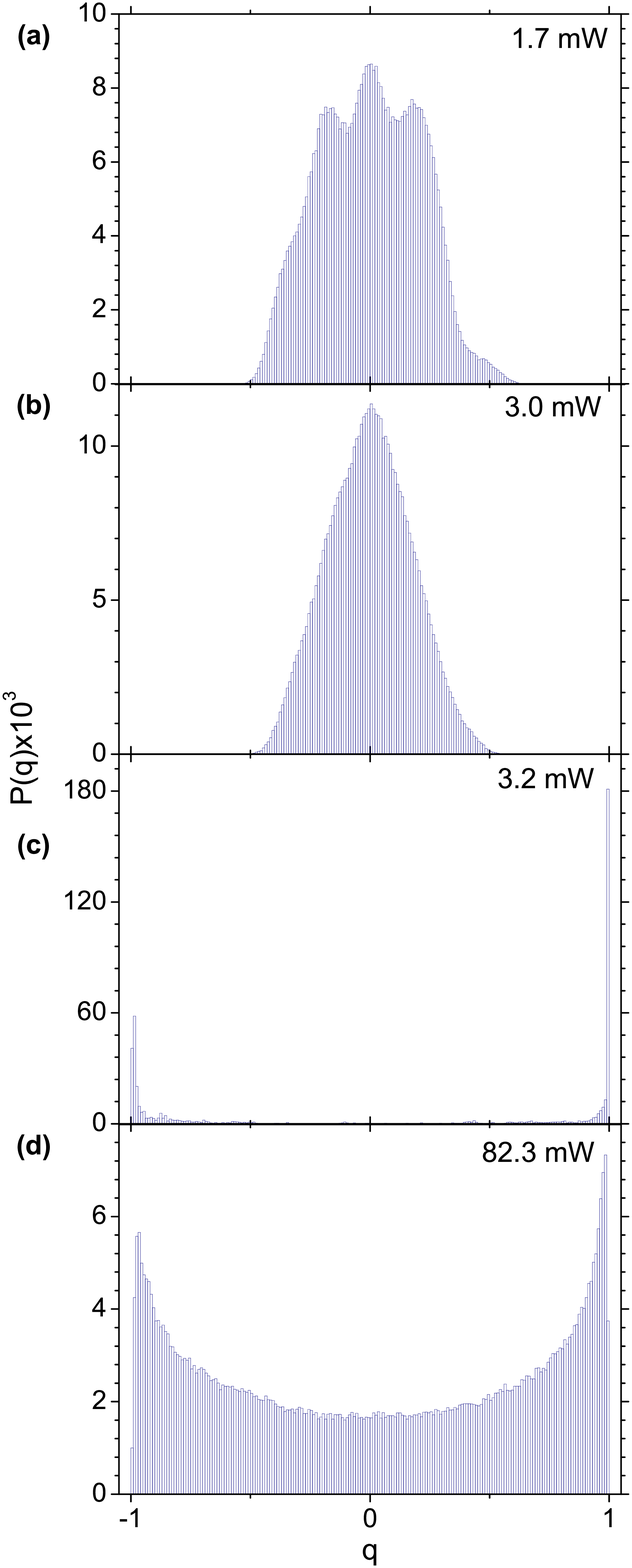}}
%%%\centerline{\epsfig{width= 8 cm,clip,figure=Fig.2_pb_v1.eps}}
%%\centerline{\epsfig{width= 7.8 cm,clip,figure=Fig.2v2.eps}}
\hspace{0.68 cm}
\epsfig{width= 7.45 cm,clip,figure=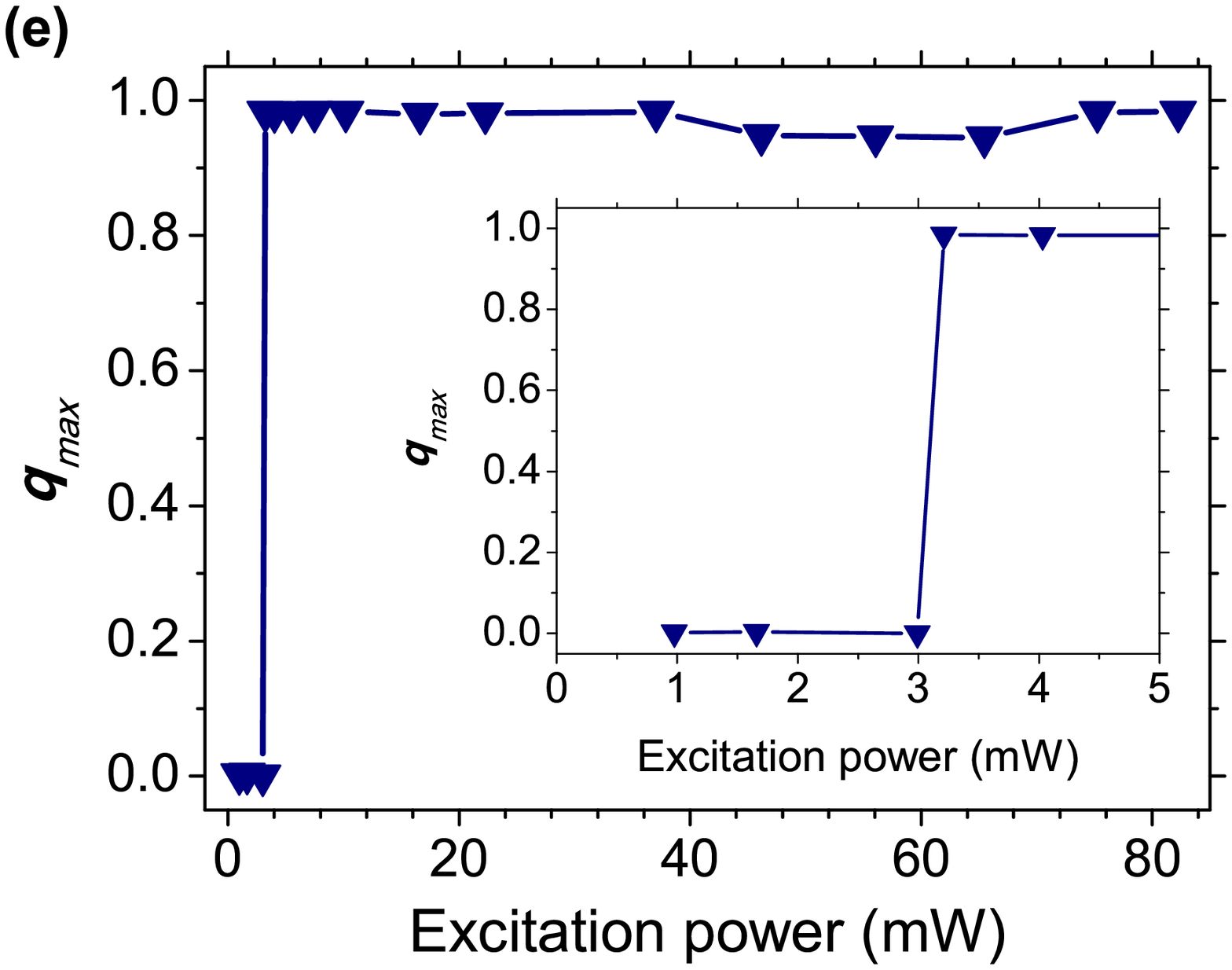}
%\vspace{-0.5cm}
\caption{%{\bf Pulse-to-pulse intensity fluctuations and corresponding overlap distributions signalizing the photonic
%RSB glassy transition in a RL system in the incoherent regime.}
%
(color online).
(a)-(d) PDF of the overlap parameter corresponding to the powers below (a,b), just above (c) and well above (d) the threshold.
(e) Dependence of the Parisi overlap order parameter $q_{\mbox{\scriptsize max}}$ on the pump power.
The inset details the threshold (solid line is a guide to the eye).
}
\label{fig2}
\end{figure}

%%FIG. 2

Figures 2(a) to 2(d) show the results for the PDF $P(q)$ obtained from the experimental data, and Fig.~2(e) shows the value $|q| = q_{\mbox{\scriptsize max}}$ at which $P(q)$ has its maximum, defining the so-called Parisi overlap parameter. Both results are in quite good agreement with the theoretical predictions and the experimental results of Ref.~\cite{22}.
A sharp transition coinciding with the threshold, $\approx 3$ mW, is observed from the photonic paramagnetic [Figs.~2(a) and (b), and excitation power $\lesssim 3$ mW in Fig.~2(e)] to
the photonic spin-glass phase with RSB [Figs.~2(c) and (d), and excitation power $\gtrsim 3$ mW in Fig.~2(e)].

The theoretical background that accounts for the present findings can be described as follows. In a series of remarkable works~\cite{24,25,26,27,28} a phase diagram for RLs with disordered nonlinear medium has been recently built based on Langevin equations for the complex amplitudes of the normal modes $a_j (t)$. For open cavity systems, the general effective Hamiltonian~\cite{27,28} includes a sum of quadratic [${\cal O} (a_j a_k^*)$] and quartic [${\cal O} (a_j a_k^* a_l a_m^*)$] disorder terms in the mode amplitudes. The physical origin of the quadratic coupling lies in the spatially inhomogeneous refractive index, as well as in a nonuniform distribution of the gain and an effective damping contribution due to the open cavity leakage. On the other hand, the quartic coupling is associated to a modulation of the nonlinear $\chi^{(3)}$ susceptibility with a random spatial profile. Such ingredients are also present in the 1D Er-RFL system analyzed in this work.

As the spatial disorder generally makes the explicit calculation of the quadratic and quartic couplings rather difficult, they have been considered in Refs.~\cite{24,25,26,27,28} as quenched Gaussian variables. The resulting photonic Hamiltonian for open-cavity systems thus becomes an analogue to that of the spherical $p$-spin model~\cite{30}, with a sum of $p = 2$ and $p = 4$ terms and Gaussian couplings. A replica-trick approach identified a phase diagram for the pumping rate as a function of the disorder strength, displaying the presence of photonic paramagnetic, ferromagnetic, and RSB spin-glass phases, depending on the trend of the disorder to hamper the synchronous oscillation of the modes~\cite{24,25,26,27,28}.

Although the disordered nonlinear medium considered in Refs.~\cite{24,25,26,27,28} is 3D, we argue that qualitatively similar features of the photonic phase diagram can be also expected to hold in lower (1D and 2D) dimensional nonlinear systems with disorder, as in the present case of the 1D Er-RFL. Indeed, we recall that the first experimental observation of a photonic RSB spin-glass phase actually took place in a 2D RL system~\cite{22}. In the herein described 1D Er-RFL system, instead of the presence of random scatterer particles, the disorder is due to the continuous fiber Bragg grating in which a random distribution of phase errors was written. As mentioned, the density of ``scatterers" in the fiber has been tuned so as to provide RL operation in the coherent feedback regime. In addition, by taking the random couplings as Gaussian variables in the photonic Hamiltonian of Refs.~\cite{24,25,26,27,28}, the explicit connection with the actual 3D structure of the disordered nonlinear medium is lost, and conceivable theoretical extensions to include other sources of disorder are actually made possible. This reasoning is reinforced by the fact that, while the summations in the magnetic $p$-spin Hamiltonian run over the spins positions in the lattice (which necessarily takes into account its explicit geometrical structure), the sums in the photonic Hamiltonian are over the mode labels, which keep no structural link with the background medium.

In conclusion, we have demonstrated the first observation of RSB in a 1D RFL system. The signature depicted by the Parisi overlap order parameter clearly identifies the photonic paramagnetic and photonic spin-glass regimes.  Our 1D results show that this transition undoubtedly coincides with the RL threshold. As the Er-RFL operates in the coherent feedback regime, due to the scattering regime leading to 1D Anderson localization in the RFL,  the glassy behavior of this RFL holds for the highest pump power available in our system. The demonstration of the RSB in the cw regime also opens up important possibilities for new experimental demonstrations of other expected phase transitions and photonic regimes~\cite{24,25,26,27,28}.

We acknowledge financial support from the Brazilian Agencies: Conselho Nacional de Desenvolvimento Cient\'{\i}fico e Tecnol\'ogico (CNPq) and Funda\c{c}\~ao de Amparo \`a Ci\^encia e Tecnologia do Estado de Pernambuco (FACEPE). The work was performed in the framework of the National Institute of Photonics (INCT de Fot\^onica) and PRONEX-CNPq/FACEPE projects. ALM acknowledges the CNPq for a postdoctoral fellowship. RK acknowledges support from the Canada Research Chairs program of the Govt. of Canada.

\bibliographystyle{unsrt}

%\clearpage

\end{document}